# Robustness study of noisy annotation in deep learning based medical image segmentation

**Shaode Yu, Erlei Zhang, Junjie Wu, Hang Yu, Zi Yang, Lin Ma, Mingli Chen, Xuejun Gu\*, Weiguo Lu**[*]

Medical Artificial Intelligence and Automation Laboratory, Department of Radiation Oncology, University of Texas Southwestern Medical Center, Dallas, TX, 75390, USA

\*E-mails: Xuejun.Gu@utsouthwestern.edu ; Weiguo.Lu@utsouthwestern.edu

**Abstract**

Partly due to the use of exhaustive-annotated data, deep networks have achieved impressive performance on medical image segmentation. Medical imaging data paired with noisy annotation are, however, ubiquitous, but little is known about the effect of noisy annotation on deep learning-based medical image segmentation. We studied the effects of noisy annotation in the context of mandible segmentation from CT images. First, 202 images of Head and Neck cancer patients were collected from our clinical database, where the organs-at-risk were annotated by one of 12 planning dosimetrists. The mandibles were roughly annotated as the planning avoiding structure. Then, mandible labels were checked and corrected by a physician to get clean annotations. At last, by varying the ratios of noisy labels in the training data, deep learning-based segmentation models were trained, one for each ratio. In general, a deep network trained with noisy labels had worse segmentation results than that trained with clean labels, and fewer noisy labels led to better segmentation. When using 20% or less noisy cases for training, no significant difference was found on the prediction performance between the models trained by noisy or clean. This study suggests that deep learning-based medical image segmentation is robust to noisy annotations to some extent. It also highlights the importance of labeling quality in deep learning.

Key words: Deep learning; Noisy annotation; Radiation oncology; Medical image segmentation

## 1. Introduction

Deep supervised networks have achieved impressive performance in medical image segmentation partly due to the use of high-quality exhaustive-annotated data (Hesamian *et al.*, 2019; Chen *et al.*, 2019; Liu *et al.*, 2017). However, in radiation oncology, it is hard or impossible to conduct sufficient high-quality image annotation. Besides potential hurdles of funding acquisitions, time cost and patient privacy, accurate annotation of medical images always requires scarce and expensive medical expertise (Greenspan *et al.*, 2016) and thereby, medical imaging data paired with noisy annotation is prevalent, particularly in radiation oncology.

An increasing attention has been paid to the issue of label noise in deeply supervised image classification (Han *et al.*, 2019; Hendrycks *et al.*, 2018; Tanaka *et al.*, 2018). These approaches to tackle the label noise could be generally categorized into two groups. One tends to analyze the label noise and to develop deep networks with noise-robust loss functions. Reed *et al* proposed a generic way to tackle inaccurate labels by augmenting the prediction objective function with a notion of perceptual consistency (Reed *et al.*, 2014). The consistency was defined as the confidence of predicted labels between different objective estimation computed from the same input data. Further, the authors introduced a convex combination of the known labels and predicted labels as the training target in self-learning. Patrini *et al* presented two procedures for loss function correction, and both the application domain and the network architecture were unknown (Patrini *et al.*, 2017). The computing cost is at most a matrix inversion and multiplication. Both procedures were proven to be robust to the noisy data, and importantly, the Hessian of the loss function was found independent from label noise for the ReLU networks. By generalizing the categorical cross entropy, Zhang and Sabuncu developed a theoretically grounded set of noise-robust loss functions (Zhang and Sabuncu, 2018). These functions could be embedded into any deep networks to yield good performance in a wide range of noisy label scenarios. And notably, Luo *et al* designed a variance regularization term to penalize the Jacobian norm of a deep network on the whole training set (Luo *et al.*, 2019). Both theoretically deduced and experimental results showed that the regularization term can decrease the subspace dimensionality, improve the robustness, and generalize well to label noise. However, these approaches require prior knowledge or an accurate estimation of the label noise distribution, which is not practical in real-world applications. The other group tends to figure out and to remove or correct noisy labels by using a small set of clean data. Misra *et al* demonstrated that noisy labels from human-centric annotation are statistically dependent on the data, and thus, clean labels could be learnt to decouple this kind of human reporting bias and to improve image captioning performance (Misra *et al.*, 2016). Xiao *et al* introduced a general framework to train deep networks with a limited number of clean samples and massive noisy samples (Xiao *et al.*, 2015). The relationships among images, class labels, and label noises were quantified with a probabilistic graphical model, which was further integrated into an end-to-end deep learning system. Mirikharaji *et al* proposed a practical framework to learn from a limited number of clean samples in the training phase that assigned higher weights to pixels with gradient directions closer to those of clean data in a meta-learning approach (Mirikharaji *et al.*, 2019). This kind of approaches is feasible but significantly increases the computing complexity.

Many efforts have been made to tackle label noise in image classification, while little is known about the effect of annotation quality on object segmentation. Object segmentation can be viewed as pixel-wise image classification and requires high-quality exhaustive-annotated data for algorithm training. However, in radiation oncology, some organs-at-risk (OARs) may be roughly annotated due to the trade-off between time spent and radiation treatment planning quality. Such rough annotations can mislead deep network training and result in ambiguous localization of anatomical structures. This study concerns the effect of annotation quality on medical image segmentation. It involves medical image data annotated by dosimetrists in radiation treatment planning and differs from the aforementioned studies, which artificially generate noisy labels and do not reflect real-life scenarios. The primary purpose of this study is to investigate whether a deep network trained with noisy data can achieve comparative performance as that trained with clean data. Specifically, the effect of different ratios of noisy cases in the training data is investigated in the context of deep learning based mandible image segmentation.

## 2. Methods and Materials

*2.1 Data Collection and Pre-processing*

A total of 202 computerized tomography (CT) images of Head and Neck cancer patients were collected (47 females and 155 males; age, 62±12 years). These CT images were exported from ARIA® Radiation Therapy Management System (ARIA RTM, 25 patients) and ARIA® Oncology Information System for Radiation Oncology (ARIA OIS RO, 177 patients) of Varian Medical Systems. The in-plane voxel resolution was between 1.17 and 1.37 mm, and the slice thickness was 3.00 mm. The in-plane image size was [512, 512] and the slice number was in the range of 127 to 264.

To reduce computing complexity, the original volumes were cropped to cover the whole mandible regions, and the voxel sizes were interpolated to the same resolution. Then, the matrix size of these volumetric images became [256, 256, 128], and their voxel size became [1.17, 1.17, 3.00] mm$^3$. In addition, to highlight the bone regions, the intensity $V$ was normalized by the mean $\mu_{V>900}$ and standard deviation $\sigma_{V>900}$ of the CT image: $V' = (V - \mu_{V>900})/\sigma_{V>900}$.

*2.2 Mandible Annotation and Correction*

Mandible is one of the OARs in 3D-conformal radiotherapy or intensity-modulated radiotherapy for head and neck cancers. The mandible regions are outlined in radiation treatment planning to avoid the potential development of jaw osteoradionecrosis. However, due to rough annotation, the labels are noisy, such as incomplete or over-segmented.

A total of 12 dosimetrists participated in CT image annotation. Among them, 3 dosimetrists each annotated more than 30 cases (31, 59, and 82 cases), and the others each annotated fewer than 10 cases. To address the issue of inaccurate annotation, the correction procedure consisting of deleting and adding operations was applied by a physician. Given the clean labels, Figure 1 shows the distribution of voxel numbers in the mandible regions, kept regions, deleted regions, and added regions with correction. Statistically, with this data set, the mandible region contained 14097±3236 voxels (≈57.89 ±13.29 cm$^3$); in the correction procedure, 95.04% voxels (13398±3474) were kept from the noisy labels on average; and to form the clean labels, 3887±3254 and 699±765 voxels are removed from and added to the noisy labels, respectively.

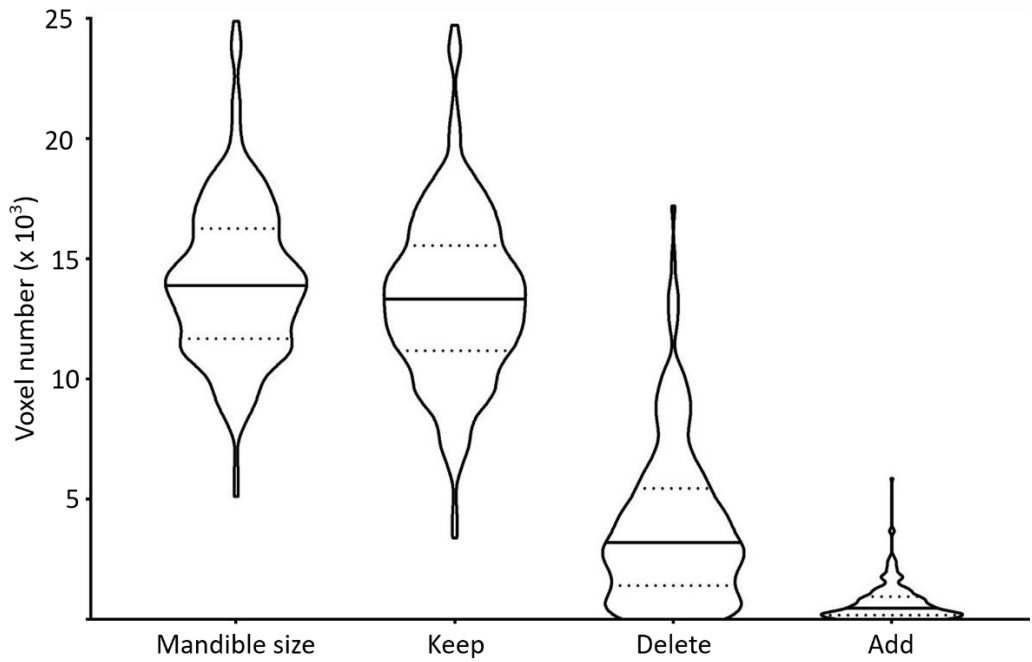

Figure. 1. Voxel number analysis of the mandible regions with median (solid line) and quartiles (dotted line) in violin plots. It indicates the average size of the mandible region. Given the clean annotation, this figure also shows the distributions of the number of voxels kept in, deleted from, and added to the noisy labels in the correction procedure.

Figure 2 further illustrates three representative examples of mandible annotation before and after correction. In the figure, the top row shows the delineation of mandible regions in radiation treatment planning and the bottom row shows the corresponding mandible regions after correction. Specifically, red and green contours stand for the boundaries of mandible regions. In general, label noise of mandible comes from incomplete annotation ((a) vs (d)), different definition of mandible ((b) with vs (e) without tooth regions), and inaccurate delineation ((c) vs (f)).

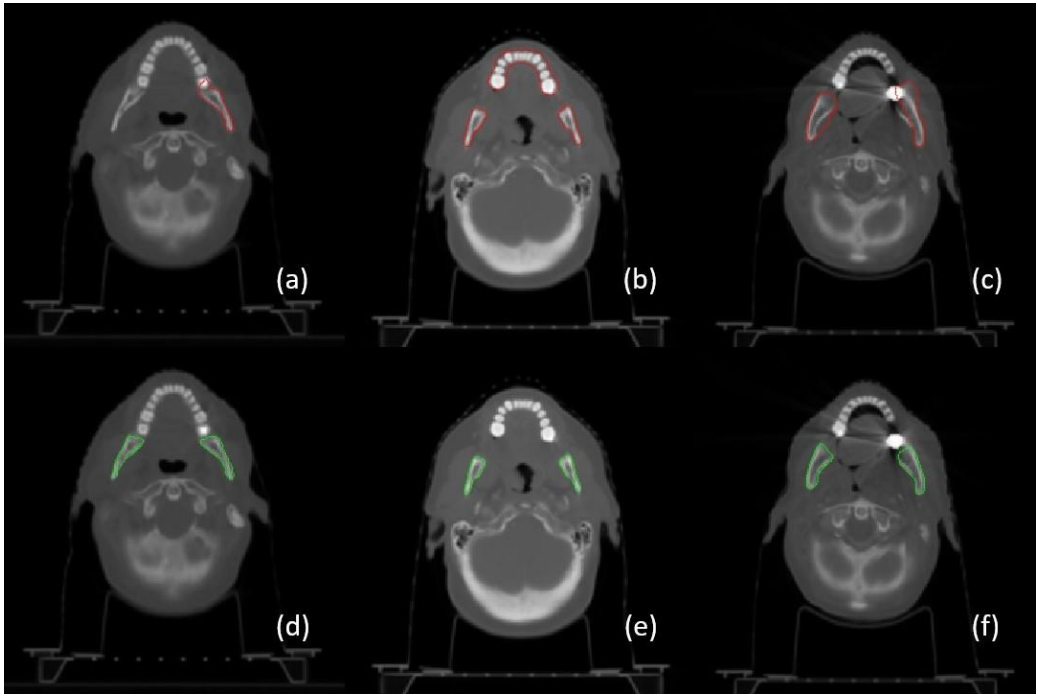

Figure. 2. Perceived annotation difference via contour visualization. The top row shows rough delineation of mandible regions in radiation treatment planning and the bottom row shows the corresponding mandible regions after label correction. It shows the label noise of mandible mainly comes from incomplete annotation ((a) vs (d)), different definition of mandible regions with tooth (b) or without tooth (e) regions and inaccurate delineation ((c) vs (f)).

## 2.3 Experiment design

The experiment design is illustrated in Figure 3. The 202 mandible image samples were divided into a training set (180 samples), a validation set (10 samples), and a testing set (12 samples). For fair comparison, samples in both the validation and the testing set are fixed. A total of 8 experiments were conducted. In the figure, a rectangle stands for 18 samples, *i.e.*, 10% training samples, the brown color stands for noisy annotation, and the green for clean labels. In each experiment, a deep network was first trained, then validated for model selection, and at last tested. The performance of deep models trained on different percentages of noisy cases was compared. Specifically, the percentages were 0% to 60% at 10% increment and 100%, labeled alphabetically by (A) to (H). In addition, the report of each percentage was based on the average performance from six random selections of noisy labels in the training data.

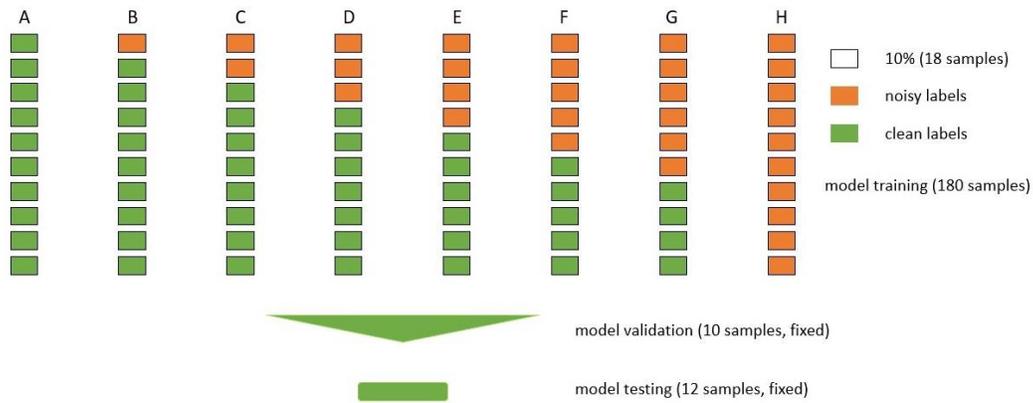

Figure. 3. Experimental design of different percentages of noisy cases on medical image segmentation. In each experiment, a deep network is trained, then validated for the selection of a well-trained model and at last tested. A total of 8 experiments are conducted and the percentages of noisy labels are 0% (A), 10% (B), 20% (C), 30% (D), 40% (E), 50% (F), 60% (G) and 100% (H). Each experiment repeats 6 times based on randomly splitting data samples in the training set and the average performance is reported.

*2.4 A deep neural network and its parameter settings*

A compact 3D convolutional network HighRes3DNet is used, which utilizes efficient and flexible elements, such as dilation convolution and residual connection, for volumetric image segmentation (Li *et al.*, 2017). It has been applied for brain parcellation and hyperintensity isolation (Kuijf *et al.*, 2019; Li *et al.*, 2017). In this study, its parameters were set as follows. Both the input and output image size were [256, 256, 128], and the output images were in binary values. The Adam optimizer was used for hyper-parameter optimization (Kingma and Ba, 2014). Binary cross entropy was set as the loss function, and ReLu as the activation function. The learning rate was $10^{-4}$, the number of iteration was $10^5$, and the batch size is 1. For each volumetric image, 64 patches of size [96, 96, 96] were uniformly generated. Other parameters were set as default with random initialization. Specifically, the model was validated per $10^3$ iterations at the training stage, and thus $10^2$ checkpoints were saved after training. Neither fine-tuning nor data augmentation was used.

*2.5 Model training, validation and selection*

Figure 4 shows an example of the model training (blue line) and validation (red circles) based on 100% of noisy cases. In the training stage, the network inferred the data samples in the validation set per $10^3$ iterations, and thus, $10^2$ check points (red circles) of loss values were obtained after the training procedure. To select the best trained model, the validation losses were compared, and the check point with the least loss value indicated the model at this point was optimized. The selected model was used for further mandible CT image segmentation. In this example, the model validated at $9.6 \times 10^4$ iterations achieved the least loss value ($8.0 \times 10^{-5}$), and it was taken as the best trained.

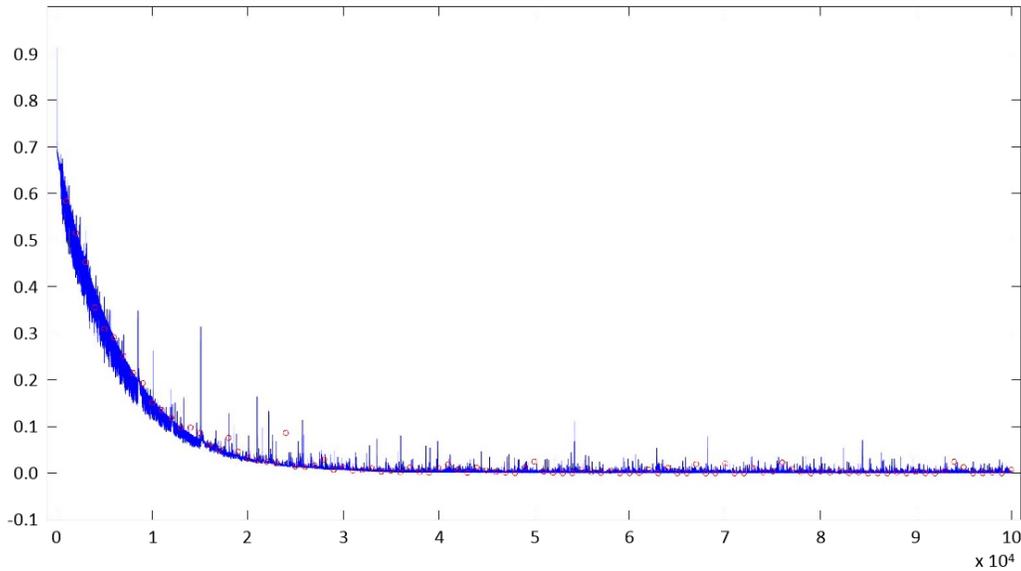

Figure. 4. Model training (blue line), validation (red circles) and selection. A model under training will be validated per $10^3$ iterations and subsequently, $10^2$ check points (red circles) are saved. To select an optimal model, the validation loss is calculated and the check point with the least loss value is selected as the optimized model for further mandible CT image segmentation.

*2.6 Performance evaluation*

Two metrics were used to verify the performance of the well-trained model on mandible segmentation. One was the Dice coefficient (DICE), which estimates the overlapping ratio between the predicted image and ground truth image. Since this study concerns a binary image segmentation problem, the mandible or predicted mandible regions were marked with value 1 and the rest with 0. A higher DICE value indicates a better segmentation. The other metric was voxel-wise false positive rate (vs-FPR), which is the number of incorrect positive predictions over the total number of negative voxels. The best vs-FPR is 0.0, and it means non-mandible regions are correctly predicted.

*2.7 Software and platform*

Experiments were conducted on a 64-bit Windows 10 workstation with 8 Intel (R) Xeon (R) processors (3.60 GHz), 64.0 GB RAM, and a NVIDIA GeForce RTX 2018 TI GPU card. HighRes3DNet was implemented in NiftyNet (Gibson *et al.*, 2018)(version 0.5.0, https://niftynet.io/), a deep learning toolbox for medical image analysis based on Tensorflow (version 1.13.2, https://www.tensorflow.org/).

## 3   Results

*3.1 The performance of selected models on the training set*

The performance of selected models on the testing set is shown in Figure 5 with error bars for the mean values of DICE and vs-FPR with respect to different percentages of noisy cases. Overall, the DICE values decreased and vs-FPR values increased as the number of noisy cases increased. It suggests that the performance of the deep network trained with noisy labels was inferior to that trained with clean labels, and more clean labels led to better predication. However, there was no significant difference in prediction performance between the model trained with clean samples and the models trained with 10% or 20% noisy cases (two sample *t*-test, *p*>0.25).

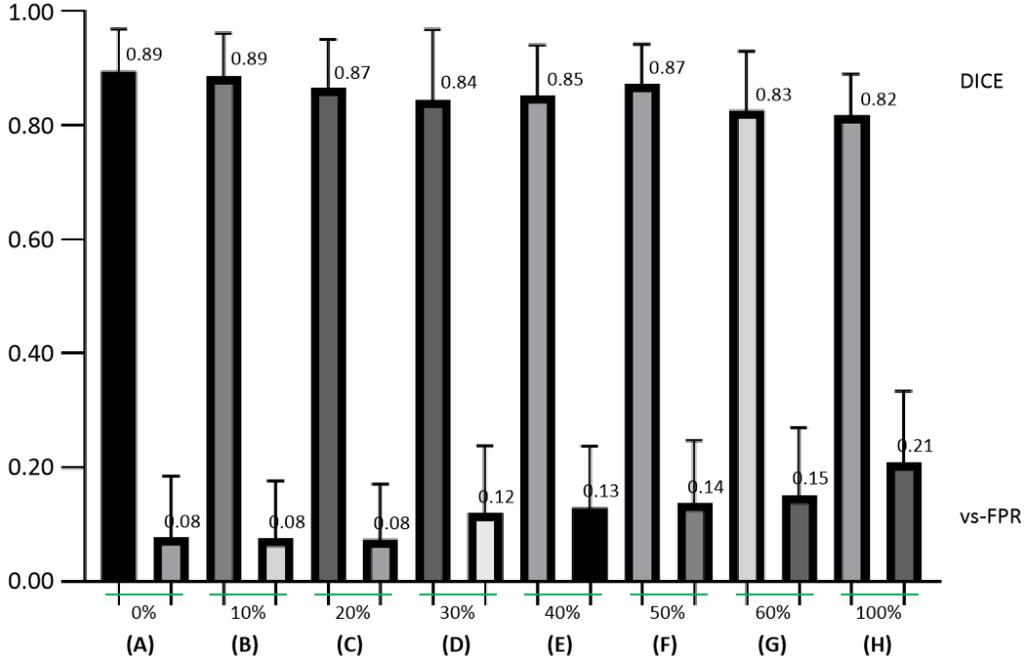

Figure. 5. The performance of selected models on the testing set. It shows the mean metric values in terms of different percentages of noisy cases. It indicates that when the number of noisy cases increases, there is a decrease trend of DICE values and an increase trend of vs-FPR values. When 20% or 10% noisy cases are used for training, there is no significant difference of metric values between the model trained with clean samples and the models trained with noisy cases (two sample *t*-test, *p*>0.25).

*3.2 The performance of selected models on the training samples*

Figure 6 shows the performance of optimized models evaluated on the training set, which quantifies the prediction results at the training stage with the model of minimum training loss. The result indicates that smaller percentages of noisy cases did not correspond to better training performance (10% noisy labels, DICE = 0.80±0.09, vs-FPR 0.08±0.09; 100% noisy labels, DICE = 0.85±0.09, vs-FPR 0.08±0.11). Interestingly, the vs-FPR remained stable ($\approx$ 8%) despite different numbers of noisy cases.

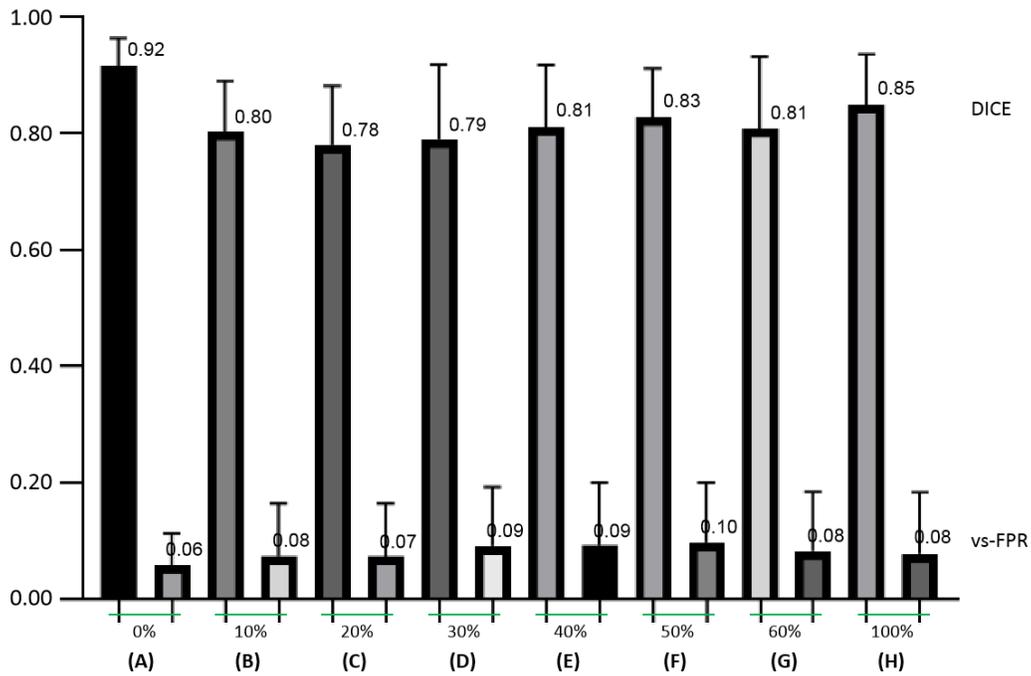

Figure. 6. The performance of selected models on the training set when using the corresponding noisy labels as the "ground truth". It shows mean values of metrics in terms of different percentages of noisy cases. It is observed less percentages of noisy cases do not correspond to better performance and the vs-FPR keeps a stable value on average despite of different numbers of noisy cases in the training stage.

When using the clean labels as the ground truth to quantify the prediction results at the training stage, the performance of selected models decreased as the number of noisy labels increased as shown in Figure 7. In detail, the DICE values decreased from 0.92±0.05 to 0.82±0.08, and the vs-FPR values increased from 0.06±0.05 to 0.19±0.10. Two sample *t*-test indicates that there was significant difference in the DICE values between the model trained with 0% noisy cases and the model trained with 20% or 10% noisy cases ($p<0.05$), but not in the vs-FPR values ($p>0.30$).

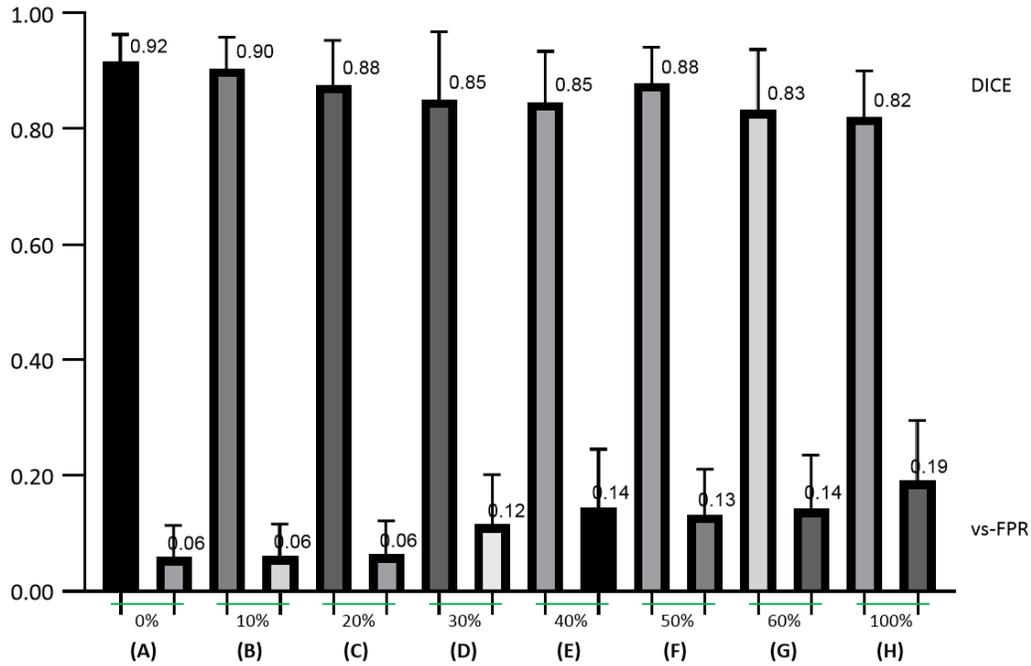

Figure. 7. The performance of selected models on the training set when using the clean labels as the ground truth. It is the results of optimized models at the iteration when the loss function achieves the least values in the model training stage. It shows mean metric values in terms of different numbers of noisy cases for training.

## 4   Discussion

This study concerns deep learning with noisy annotation and explores mandible image segmentation for radiation treatment planning. It demonstrates the prediction performance of deep networks trained with different number of noisy cases, suggesting that deep learning is robust to noisy annotation. The performance of mandible segmentation at the training stage was compared for using noisy labels and clean labels as the ground truth. It indicates the entropy-based training loss is a good driver for the segmentation task regardless of the training data quality when the same data is used for both training and testing. It also suggests that the prediction model may improve with bootstrap ensemble approaches since training with clean labels results in most consistent high-quality performance.

### *4.1 The quality of dosimetrists' annotation*

The initial image annotation by 12 dosimetrists with different experience showed inter-rater disagreement in mandible, such as the inclusion of teeth regions. Mandible is one of OARs delineated with less care, and the annotation may be incomplete or over-segmented (Figure 2). The initial annotation quality, however, was acceptable since most voxels were kept in the correction procedure (Figure 1). Given the clean labels, the quality of initial annotation can be estimated with DICE and vs-FPR as shown in Figure 8. It shows that the median value of DICE and vs-FPR were 0.87 and 0.19, respectively, and 45 out of 202 cases had DICE < 0.80 and vs-FPR > 0.20. On average, the DICE was 0.87±0.08, and the vs-FPR was 0.20±0.13.

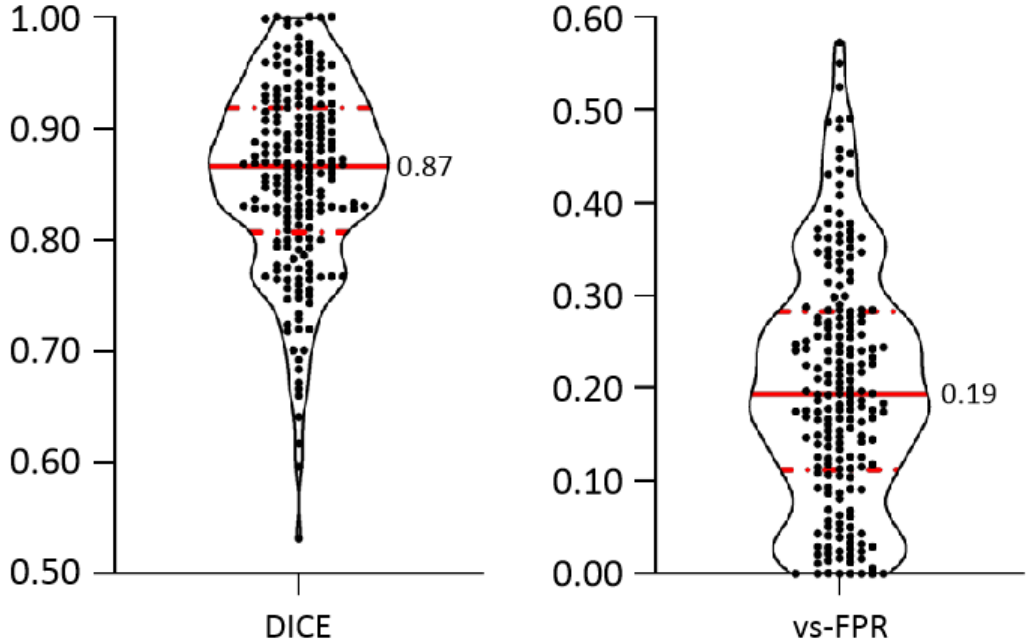

Figure. 8. The quality of dosimetrists' annotation in radiation treatment planning. Given the clean labels, annotation quality is estimated with DICE and vs-FPR. The violin plots show the median (red solid lines) and quartiles (red dashed lines).

The noisy annotation based mandible segmentation distinguishes this study from those noisy labeling based image classification studies. In mandible segmentation, the number of noisy cases can be controlled as 10% or 20% as that in image classification studies. However, it is hard to guarantee the annotation quality of each case which imposes more difficulties on the simulations of deep learning with noisy annotation. Moreover, in aforementioned image classification studies, hundreds of thousands data samples could be collected and in particular, incorrect labels could be synthetically generated by following a uniform or class-dependent asymmetric noise distribution, while in the community of radiation oncology, data collection and exhaustive annotation are always challenging. Specifically, manual annotation of CT images for treatment planning is a real-world scenario in radiation oncology and thus, deep learning with noisy annotation should be paid extra attention due to its wide applications.

*4.2 The robustness of deep learning to noisy annotation*

It is worth noting that deep learning is robust to noisy annotation to some extent. Based on noisy cases (Figure 8, preliminary annotation quality, DICE 0.87±0.08 and vs-FPR 0.20±0.13), this study shows that the selected model was well-trained (Figure 7, DICE 0.82±0.08 and vs-FPR 0.19±0.10) and obtained good segmentation on the testing set (Figure 5, DICE 0.82±0.07 and vs-FPR 0.21±0.13). In particular, there was no significant difference in prediction performance on the testing set between the model trained with only clean cases and the models trained with 10% or 20% noisy cases (Figure 5, two sample *t*-test, $p>0.25$). A similar phenomenon has been observed in image classification. Van Horn *et al* claimed that if the training set is sufficiently large, a small label error rate of training data led to an acceptable small increase of prediction error in the test set (Van Horn *et al.*, 2015). Rolnick *et al* observed that deep learning was robust to label noise, and they figured out that a sufficient training set can accommodate a wide range of noise levels (Rolnick *et al.*, 2017). This kind of robustness to label noise can be explained from the capacity of deep networks on continual representation learning. Rolnick *et al* also pointed out that a deep architecture tends to learn intrinsic pattern of objects instead of merely memorizing noise (Rolnick *et al.*, 2017). Arpit *et al* conducted close examinations at the memorization of deep networks with regard to the model capacity, generalization, and adversarial robustness (Arpit *et al.*, 2017). They showed that deep networks preferred to prioritize learning simple and general patterns before fitting the noise, which may explain the limited vs-FPR values in the training stage (Figure 6).

*4.3 The importance of annotation quality in deep learning*

The importance of labeling quality has been highlighted in previous studies, and this study further emphasizes this point. Given the annotation quality of dosimetrists (DICE, 0.87±0.08; vs-FPR, 0.20±0.13; Figure 8), results indicate that the selected model trained with 100% noisy cases achieved DICE 0.82±0.07; vs-FPR 0.21±0.13 (Figure 5), both of which were worse than the annotation quality of dosimetrists. Even in the training stage, the selected model obtained inferior performance quality (DICE 0.82±0.08; vs-FPR 0.19±0.10; Figure 7) than the initial annotation. On the contrary, when using clean data for model training, the selected model reached DICE 0.92±0.05 and vs-FPR 0.06±0.05 on the training set (Figure 7) and DICE 0.89±0.07 and vs-FPR 0.08±0.11 on the testing set (Figure 5). In particular, noisy annotation misled model training. For instance, when using 100% noisy cases for model training, it causes pseudo-high performance by comparing the DICE and vs-FPR values in Figure 6 and Figure 7.

A deep network is a computational model. It consists of multiple levels of abstraction for representation learning. Through an iterative process of forward pass and back propagation, a deep network attempts to learn an intricate structure from a large number of data samples and its internal hyper-parameters are dynamically updated toward an optimal solution (LeCun *et al.*, 2015). To deep supervised learning, any noisy annotation is bound to mislead the iterative process and subsequently, the learnt structure may not be representative, and the determined parameters may not be optimal. Thus, it is essential to provide clean annotation for deep learning based image segmentation.

*4.4 Future work on deep learning with noisy annotation*

In radiation oncology, it is possible to provide a small number of high-quality exhaustive-annotated samples and thus, how to utilize such limited samples to address the issue of deep learning with noisy annotation becomes an urgent problem.

Several studies for image classification have provided clues on this topic. One way is to implicitly estimate the possibility of training samples with clean labels. Reed *et al* proposed a coherent bootstrapping model to evaluate the consistency between given labels and its predicted labels in the training stage, and both labels contributed to the resultant prediction in a convex combination (Reed *et al.*, 2014). Han *et al* trained two deep networks simultaneously, and the networks learned from each other and mutually exchanged probable clean labels to reduce the error flows (Han *et al.*, 2018). As such, each network could attenuate different types of labeling errors and lead to better performance. The other way is to explicitly guide the model training with clean labels. Given clean samples in the validation set, Ren *et al* designed an online meta-learning algorithm, which updated the weights of training examples using a gradient descent step on the current training example weights to minimize the loss on a validation set (Ren *et al.*, 2018). Mirikharaji *et al* developed an adaptive reweighting approach and commensurately treated both clean and noisy annotations in the loss function (Mirikharaji *et al.*, 2019). They deployed a meta-learning approach to assign higher importance to pixels whose loss gradient direction was closer to those of clean data. Besides, utilizing a noise-robust loss function could further improve the training effectiveness (Zhang and Sabuncu, 2018).

The optimization of deep networks is content-aware and the purpose is to retrieve patterns shared by training samples (LeCun *et al.*, 2015). Thus, building multiple atlases for a specific application potentially improves the representation learning performance as reported in the literature. Ma *et al* utilized deep learning to localize the prostate region and to distinguish prostate pixels from the surround tissues (Ma *et al.*, 2017), and used similar atlases to refine the segmentation results. Zhu *et al* proposed a hybrid framework for the fusion of predicted hippocampus regions (Zhu *et al.*, 2020). The framework first used atlases to estimate the deformation of image labels, and then a fully convolutional network was designed to learn the relationship between pairs of image patches to correct the potential errors. Finally, both multi-atlas image segmentation and the fully convolutional network were used for label fusion. Vakalopoulou *et al* introduced a multi-network architecture to exploit domain knowledge (Vakalopoulou *et al.*, 2018). After co-aligning multiple anatomies through multi-metric non-rigid registration, each network performed CT image segmentation for interstitial lung disease in the atlas space. At last, segmentation results were fused in the source data space. Ding *et al* presented a deep learning based label fusion strategy (Ding *et al.*, 2019). It attempted to locally select a set of reliable atlases by deep learning, and finally, the estimated labels were fused via plurality voting.

Deep learning with noisy annotation is challenging but clinically important (Min *et al.*, 2019; Rozario *et al.*, 2017; Sahiner *et al.*, 2019; Yang *et al.*, 2019). Deep learning shows robustness to noisy annotation, while for further improvement, it should take advantage of clean data samples either implicitly or explicitly to avoid the misleading of noisy annotation in the training stage. Besides, multiple atlases provide diverse but representative contextures and multi-atlas image segmentation provides good spatial consistency via deformable segmentation, both of which might contribute to the development of deep networks robustness and may be generalizable to noisy annotation.

## 5. Conclusions

This study concerns deep learning with noisy annotation in medical image segmentation. It shows deep learning is robust to noisy annotation to some extent. In general, a deep network trained with noisy labels is inferior to that trained with clean labels. Thus, how to utilize limited clean samples to improve the performance of deep learning with noisy annotation should be paid extra attention.

**References**

Arpit D, Jastrzębski S, Ballas N, Krueger D, Bengio E, Kanwal M S, Maharaj T, Fischer A, Courville A and Bengio Y 2017 A closer look at memorization in deep networks *Proceedings of the 34th International Conference on Machine Learning* 233-42

Chen H, Lu W, Chen M, Zhou L, Timmerman R, Tu D, Nedzi L, Wardak Z, Jiang S and Zhen X 2019 A recursive ensemble organ segmentation (REOS) framework: Application in brain radiotherapy *Physics in Medicine & Biology* **64** 025015

Ding Z, Han X and Niethammer M 2019 VoteNet: A deep learning label fusion method for multi-atlas segmentation *International Conference on Medical Image Computing and Computer-Assisted Intervention* 202-10

Gibson E, Li W, Sudre C, Fidon L, Shakir D I, Wang G, Eaton-Rosen Z, Gray R, Doel T and Hu Y 2018 NiftyNet: a deep-learning platform for medical imaging *Computer methods and programs in biomedicine* **158** 113-22

Greenspan H, Van Ginneken B and Summers R M 2016 Guest editorial deep learning in medical imaging: Overview and future promise of an exciting new technique *IEEE Transactions on Medical Imaging* **35** 1153-9

Han B, Yao Q, Yu X, Niu G, Xu M, Hu W, Tsang I and Sugiyama M 2018 Co-teaching: Robust training of deep neural networks with extremely noisy labels *Advances in Neural Information Processing Systems* 8527-37

Han J, Luo P and Wang X 2019 Deep self-learning from noisy labels *Proceedings of the IEEE International Conference on Computer Vision* 5138-47

Hendrycks D, Mazeika M, Wilson D and Gimpel K 2018 Using trusted data to train deep networks on labels corrupted by severe noise *Advances in Neural Information Processing Systems* 10456-65

Hesamian M H, Jia W, He X and Kennedy P 2019 Deep learning techniques for medical image segmentation: Achievements and challenges *Journal of digital imaging* **32** 582-96

Kingma D P and Ba J 2014 Adam: A method for stochastic optimization *arXiv preprint arXiv:1412.6980*

Kuijf H J, Biesbroek J M, De Bresser J, Heinen R, Andermatt S, Bento M, Berseth M, Belyaev M, Cardoso M J and Casamitjana A 2019 Standardized assessment of automatic segmentation of white matter hyperintensities and results of the WMH segmentation challenge *IEEE Transactions on Medical Imaging* **38** 2556-68

LeCun Y, Bengio Y and Hinton G 2015 Deep learning *Nature* **521** 436-44

Li W, Wang G, Fidon L, Ourselin S, Cardoso M J and Vercauteren T 2017 On the compactness, efficiency, and representation of 3D convolutional networks: brain parcellation as a pretext task *International Conference on Information Processing in Medical Imaging* 348-60

Liu Y, Stojadinovic S, Hrycushko B, Wardak Z, Lau S, Lu W, Yan Y, Jiang S B, Zhen X and Timmerman R 2017 A deep convolutional neural network-based automatic delineation strategy for multiple brain metastases stereotactic radiosurgery *PloS one* **12**

Luo Y, Zhu J and Pfister T 2019 A simple yet effective baseline for robust deep learning with noisy labels *arXiv preprint arXiv:1909.09338*

Ma L, Guo R, Zhang G, Tade F, Schuster D M, Nieh P, Master V and Fei B 2017 Automatic segmentation of the prostate on CT images using deep learning and multi-atlas fusion *Medical Imaging 2017: Image Processing,2017* 101332O

Min S, Chen X, Zha Z-J, Wu F and Zhang Y 2019 A two-stream mutual attention network for semi-supervised biomedical segmentation with noisy labels *Proceedings of the AAAI Conference on Artificial Intelligence* 4578-85

Mirikharaji Z, Yan Y and Hamarneh G 2019 Learning to segment skin lesions from noisy annotations *Domain Adaptation and Representation Transfer and Medical Image Learning with Less Labels and Imperfect Data* 207-15

Misra I, Lawrence Zitnick C, Mitchell M and Girshick R 2016 Seeing through the human reporting bias: Visual classifiers from noisy human-centric labels *Proceedings of the IEEE Conference on Computer Vision and Pattern Recognition* 2930-9

Patrini G, Rozza A, Krishna Menon A, Nock R and Qu L 2017 Making deep neural networks robust to label noise: A loss correction approach *Proceedings of the IEEE Conference on Computer Vision and Pattern Recognition* 1944-52

Reed S, Lee H, Anguelov D, Szegedy C, Erhan D and Rabinovich A 2014 Training deep neural networks on noisy labels with bootstrapping *arXiv preprint arXiv:1412.6596*

Ren M, Zeng W, Yang B and Urtasun R 2018 Learning to reweight examples for robust deep learning *arXiv preprint arXiv:1803.09050*

Rolnick D, Veit A, Belongie S and Shavit N 2017 Deep learning is robust to massive label noise *arXiv preprint arXiv:1705.10694*

Rozario T, Long T, Chen M, Lu W and Jiang S 2017 Towards automated patient data cleaning using deep learning: A feasibility study on the standardization of organ labeling *arXiv preprint arXiv:1801.00096*

Sahiner B, Pezeshk A, Hadjiiski L M, Wang X, Drukker K, Cha K H, Summers R M and Giger M L 2019 Deep learning in medical imaging and radiation therapy *Medical physics* **46** e1-e36

Tanaka D, Ikami D, Yamasaki T and Aizawa K 2018 Joint optimization framework for learning with noisy labels *Proceedings of the IEEE Conference on Computer Vision and Pattern Recognition* 5552-60

Vakalopoulou M, Chassagnon G, Bus N, Marini R, Zacharaki E I, Revel M-P and Paragios N 2018 AtlasNet: Multi-atlas non-linear deep networks for medical image segmentation *International Conference on Medical Image Computing and Computer-Assisted Intervention* 658-66

Van Horn G, Branson S, Farrell R, Haber S, Barry J, Ipeirotis P, Perona P and Belongie S 2015 Building a bird recognition app and large scale dataset with citizen scientists: The fine print in fine-grained dataset collection *Proceedings of the IEEE Conference on Computer Vision and Pattern Recognition* 595-604

Xiao T, Xia T, Yang Y, Huang C and Wang X 2015 Learning from massive noisy labeled data for image classification *Proceedings of the IEEE Conference on Computer Vision and Pattern Recognition* 2691-9

Yang Q, Chao H, Nguyen D and Jiang S 2019 A Novel Deep Learning Framework for Standardizing the Label of OARs in CT *Workshop on Artificial Intelligence in Radiation Therapy* 52-60

Zhang Z and Sabuncu M 2018 Generalized cross entropy loss for training deep neural networks with noisy labels *Advances in Neural Information Processing Systems* 8778-88

Zhu H, Adeli E, Shi F and Shen D 2020 FCN based label correction for multi-atlas guided organ segmentation *Neuroinformatics* 1-13